5 pages, no macros, 2 figures
\documentstyle[prl,aps]{revtex}
\begin{document}
\draft
\title{Transverse momentum and energy correlations \\
in the equilibrium system \\
from high-energy nuclear collisions
\footnote{Phys. Lett. {\bf 439} (1998) 6.}}
\author{Stanis\l aw Mr\' owczy\' nski\footnote{Electronic address:
mrow@fuw.edu.pl}}

\address{So\l tan Institute for Nuclear Studies,\\
ul. Ho\.za 69, PL - 00-681 Warsaw, Poland\\
and Institute of Physics, Pedagogical University,\\
ul. Konopnickiej 15, PL - 25-406 Kielce, Poland}

\date{26-th June 1998, revised 21-st July 1998}

\maketitle

\begin{abstract}
The so-called $\Phi$ parameter, which measures the transverse momentum
or energy correlations (fluctuations) in high-energy collisions independently 
of the particle multiplicity, is computed for the equilibrium ideal gas. As 
expected, $\Phi$ vanishes for the particles obeying Boltzmann statistics 
but is finite for the quantum particles, positive for bosons and negative 
for fermions. $\Phi_{p_{\perp}}$, which is found for the pions gas, 
significantly exceeds the value of $\Phi_{p_{\perp}}$ measured by the 
NA49 experiment. The discrepancy is discussed.

\end{abstract}

\vspace{0.5cm}
 

\vspace{0.5cm}
%\newpage

There is a variety of correlations observed in proton--proton or
proton--antiproton interactions at high energies. In particular, 
it has been found that the average particle transverse momentum depends 
on the particle multiplicity in a given collision \cite{Kaf77,Arn82}.
These correlations should be also present in nucleus--nucleus (A--A)
collisions if such a collision is a superposition of nucleon--nucleon 
(N--N) interactions. However, there is no straightforward method to observe 
them since the final state particles in A--A collisions originate from 
the various N--N interactions while the correlated particles come only 
from the same N--N interaction. 

In our earlier paper \cite{Gaz92} we have introduced a rather 
tricky quantity (later called $\Phi$) which appears to be sensitive to 
the correlations independently of the particle multiplicity. If the A--A 
collision is a superposition of N--N interactions with no secondary
collisions, the value of $\Phi$ is exactly the same for the N--N and A--A 
case. If the secondary interactions play an important role in nucleus--nucleus
collisions, the correlation of interest is reduced \cite{Ble98} and in the 
limiting case, when the final sate particles are independent from each other, 
$\Phi$ equals zero. The method developed in \cite{Gaz92} has been recently 
applied to the experimental data and it has been found \cite{Rol97,Adl98} that 
the correlation, which is present in N--N collisions, survives in 
proton--nucleus ones but is significantly reduced in the central Pb--Pb 
collisions. This result appears to be a very restrictive test of the models
describing the N--N and A--A collisions. For example, the so-called
random walk model \cite{Leo97} is ruled out because it has been shown 
to produce, in contrast to the data, the stronger correlations in A--A 
than N--N case \cite{Gaz97}.

Reduction of the correlations measured by $\Phi$ in the central A--A 
collisions is naturally associated with the evolution of the system produced 
in these collisions towards the thermodynamical equilibrium. However,
it has been correctly observed in \cite{Gaz97} that in the thermodynamical 
equilibrium the rudimentary correlation should be present. Our aim is to 
substantiate this observation.

Let us first introduce the correlation (or fluctuation) measure $\Phi_x$, 
where $x$ is a single particle characteristics such as the particle energy 
or transverse momentum. We define the variable 
$z_x \buildrel \rm def \over = x - \overline{x}$ with the overline denoting
averaging over a single particle inclusive distribution. As seen 
$\overline{z}_x = 0$. We now introduce the variable $Z$, which is 
a multiparticle analog of $z$, defined as 
$Z_x \buildrel \rm def \over = \sum_{i=1}^{N}(x_i - \overline{x})$, where the 
summation runs over particles from a given event i.e. the particles which
are produced in the collision. One observes that $\langle Z_x \rangle = 0$, 
where $\langle ... \rangle$ represents averaging over events (collisions).
Finally, we define the measure $\Phi_x$ in the following way
\begin{equation}\label{phi}
\Phi_x \buildrel \rm def \over = 
\sqrt{\langle Z_x^2 \rangle \over \langle N \rangle} -
\sqrt{\overline{z_x^2}} \;.
\end{equation}

Our purpose is to calculate $\Phi_x$ in the ideal quantum gas. At the beginning
we identify $x$ with the particle energy $E$ and then we consider the
particle transverse momentum $p_{\perp}$. 

One immediately finds that
\begin{equation}\label{phiE1}
\overline{z^2_E} = {1 \over \rho}\int{d^3p \over (2\pi )^3}
\,(E - \overline{E})^2{1\over \lambda^{-1}e^{\beta E} \pm 1} \;,
\end{equation}
where the single particle average energy is
$$
\overline E = {1 \over \rho}\int{d^3p \over (2\pi )^3} \;
{E \over \lambda^{-1}e^{\beta E} \pm 1}
$$
while $\rho$ equals
\begin{equation}\label{rho}
\rho = \int{d^3p \over (2\pi )^3} \;
{1 \over \lambda^{-1}e^{\beta E} \pm 1} \;;
\end{equation}
$\beta \equiv T^{-1}$ is the inverse temperature; 
$\lambda \equiv e^{\beta \mu}$ denotes the fugacity and $\mu$ the chemical 
potential; $E \equiv \sqrt{m^2 + {\bf p}^2}$ with $m$ being the particle
mass and ${\bf p}$ its momentum; the upper sign is for fermions while the
lower one for bosons. It is worth noting that the result (\ref{phiE1}) does 
not depend on the number of the particle internal degrees of freedom.

Since $Z_E = U - N\overline{E}$, where $U$ is the system energy,
$\langle Z_E^2 \rangle$ is computed as
$$
\langle Z_E^2 \rangle = {1 \over Xi}\, 
\Bigg[ {\partial^2 \over  \partial \beta^2}
+2 \overline{E} \, \lambda {\partial^2 \over \partial \beta \,\partial\lambda}
+\overline{E}^2\lambda {\partial \over  \partial \lambda}
\bigg(\lambda {\partial \over  \partial \lambda}\bigg) \Bigg]
\,\Xi(V,T,\lambda) \;,
$$
where $\Xi(V,T,\lambda)$ is the grand canonical partition function
\cite{Hua63} defined as
$$
\Xi(V,T,\lambda) = \sum_N \sum_{\alpha} 
\lambda^N e^{-\beta U_{\alpha}} \;,
$$
with $V$ denoting the system volume and the index $\alpha$ numerating 
the system quantum states. As well known \cite{Hua63}, the grand canonical 
partition function of the quantum ideal gas equals
$$
{\rm ln}\,\Xi(V,T,\lambda) = \pm g\,V  \int{d^3p \over (2\pi )^3} \;
{\rm ln}\big[1 \pm \lambda\,e^{-\beta E} \big] \;,
$$
with $g$ being the number of the particle internal degrees of freedom.
Consequently,
\begin{equation}\label{phiE2}
{\langle Z_E^2 \rangle \over \langle N \rangle }= 
{1 \over \rho}\int{d^3p \over (2\pi )^3}
\,(E - \overline{E})^2 {\lambda^{-1}e^{\beta E}
\over (\lambda^{-1}e^{\beta E} \pm 1)^2} \;.
\end{equation}
As previously the result is independent of $g$. One observes that 
$$
{\langle Z_E^2 \rangle \over \langle N \rangle } < \overline{z^2_E}
\;\;\;\;\;\;\; {\rm and} \;\;\;\;\;\;\; \Phi_E < 0
$$
for fermions,
$$
{\langle Z_E^2 \rangle \over \langle N \rangle }> \overline{z^2_E}
\;\;\;\;\;\;\; {\rm and} \;\;\;\;\;\;\; \Phi_E > 0
$$
for bosons and
$$
{\langle Z_E^2 \rangle \over \langle N \rangle }= \overline{z^2_E}
\;\;\;\;\;\;\; {\rm and} \;\;\;\;\;\;\; \Phi_E = 0
$$
in the classical limit i.e. when $\lambda^{-1} \gg 1$.

In the case of massless particles with vanishing chemical potential
(which corresponds to $\lambda = 1$), one finds $\Phi_E$ analytically.
Namely,
$$
\rho = {\zeta(3) \over \pi^2} {3/4 \choose 1}\, T^3
\cong {0.09 \choose 0.12} \, T^3
$$
and 
$$
\overline{E} = {\pi^4 \over 30 \zeta(3)} {7/6 \choose 1}\, T^3
\cong {3.15 \choose 2.70} \, T \;,
$$
where $\zeta(x)$ is the Riemann zeta function ($\zeta(3) \cong 1.202$,
$\zeta(5) \cong 1.037$); the upper case is for fermions and the lower one 
for bosons. Further one computes
\begin{equation}\label{phi0E1}
\overline{z^2_E} = {12 \zeta(5) \over \zeta(3)}{5/4 \choose 1}\, T^2
- \overline{E}^2 \cong {3.01 \choose 3.06} \, T^2 \;,
\end{equation}
\begin{equation}\label{phi0E2}
{\langle Z_E^2 \rangle \over \langle N \rangle }
={1 \over \rho}\Bigg[{2\pi^2 \over 15}{7/8 \choose 1}\,T^5 
- {6 \zeta(3) \over \pi^2}{3/4 \choose 1}\overline{E}\,T^4
+{1 \over 6}{1/2 \choose 1}\overline{E}^2\,T^3 \Bigg]
\cong {2.77 \choose 4.59}\,T^2 \;,
\end{equation}
which finally give
$$
\Phi_E \cong {-0.07 \choose \;\;\;0.40} \,T \;.
$$

When the system is composed of the equal mass fermions and bosons with 
the numbers of the internal degrees of freedom $g_f$ and, respectively, 
$g_b$, the analogs of the formulas (\ref{phiE1}) and (\ref{phiE2}) read
\begin{equation}\label{phifb1}
\overline{z^2_E} = {1 \over g_f\rho_f + g_b\rho_b}
\int{d^3p \over (2\pi )^3}
\,(E - \overline{E})^2 \Bigg[
{g_f \over \lambda^{-1}_f e^{\beta E} + 1} +
{g_b \over \lambda^{-1}_b e^{\beta E} - 1} \Bigg]\;,
\end{equation}
\begin{equation}\label{phifb2}
{\langle Z_E^2 \rangle \over \langle N \rangle}= 
{1 \over g_f\rho_f + g_b\rho_b}
\int{d^3p \over (2\pi )^3}
\,(E - \overline{E})^2 \Bigg[
 {g_f\lambda^{-1}_f e^{\beta E} \over (\lambda^{-1}_f e^{\beta E} + 1)^2} 
+{g_b\lambda^{-1}_b e^{\beta E} \over (\lambda^{-1}_b e^{\beta E} - 1)^2} 
\Bigg] \;.
\end{equation}

Using eqs. (\ref{phi0E1}, \ref{phi0E2}) and (\ref{phifb1}, \ref{phifb2})
one easily computes $\Phi_E$ for the baryonless quark-gluon plasma of
two flavours where $g_f = 24$ and $g_b = 16$; $\Phi_E \cong 0.17 \, T$.

Eqs. (\ref{phiE1}, \ref{phiE2}) can be used to get the measure
$\Phi_{p_{\perp}}$ of the transverse momentum fluctuations. Since 
$p_{\perp} = p \, {\rm sin}\Theta$ with $p \equiv \vert {\bf p} \vert$ and 
$\Theta$ being the angle between ${\bf p}$ and the beam $(z)$ axis
one gets
\begin{eqnarray}\label{phipt1}
\overline{z^2_{p_{\perp}}} &=& {1 \over \rho}\int{d^3p \over (2\pi )^3}
\,\Big(p_{\perp} - \overline{p}_{\perp} \Big)^2 \,
{1\over \lambda^{-1}e^{\beta E} \pm 1} 
\\[2mm] \nonumber
&=& {1 \over \rho}\int{d^3p \over (2\pi )^3}
\,\Big( {2 \over 3} p^2 - {\pi \over 2}\overline{p}_{\perp} \, p 
+ \overline{p}_{\perp}^2 \Big) \,
{1\over \lambda^{-1}e^{\beta E} \pm 1} \;,
\end{eqnarray}
\begin{eqnarray}\label{phipt2}
{\langle Z_{p_{\perp}}^2 \rangle \over \langle N \rangle }
&=& {1 \over \rho}\int{d^3p \over (2\pi )^3}
\,\Big(p_{\perp} - \overline{p}_{\perp} \Big)^2 \,
{\lambda^{-1}e^{\beta E} \over (\lambda^{-1}e^{\beta E} \pm 1)^2} 
\\[2mm] \nonumber
&=& {1 \over \rho}\int{d^3p \over (2\pi )^3}
\,\Big( {2 \over 3} p^2 - {\pi \over 2}\overline{p}_{\perp} \, p 
+ \overline{p}_{\perp}^2 \Big) \,
{\lambda^{-1}e^{\beta E} \over (\lambda^{-1}e^{\beta E} \pm 1)^2} \;,
\end{eqnarray}
where
\begin{equation}\label{pt-av}
\overline{p}_{\perp} = {1 \over \rho}\int{d^3p \over (2\pi )^3} \;
{p_{\perp} \over \lambda^{-1}e^{\beta E} \pm 1} 
= {1 \over \rho}\int{d^3p \over (2\pi )^3} \;
{p \, {\rm sin}\Theta \over \lambda^{-1}e^{\beta E} \pm 1} 
= {\pi \over 4\rho}\int{d^3p \over (2\pi )^3} \;
{p \over \lambda^{-1}e^{\beta E} \pm 1} \;.
\end{equation}

Let us observe that $\Phi_{p_{\perp}}$ is invariant under the Lorentz
boosts along the beam axis. This is evident when 
eqs.~(\ref{phipt1}, \ref{phipt2}) and (\ref{rho}, \ref{pt-av}) 
are written in the from which reveals the transformation properties. 
We consider as an example the average transverse momentum which
can be expressed as
$$
\overline{p}_{\perp} = {1 \over \rho}\int
{d^2p_{\perp} dp_{\parallel} \over (2\pi )^3} \;
{p_{\perp} \over \lambda^{-1}e^{\beta u^{\nu}p_{\nu}} \pm 1} \;,
$$
with
$$
\rho = \int
{d^2p_{\perp} dp_{\parallel} \over (2\pi )^3} \;
{1 \over \lambda^{-1}e^{\beta u^{\nu}p_{\nu}} \pm 1} \;,
$$
where $u^{\nu}$ is the four-velocity which determines the reference
frame; $u^{\nu}=(1,0,0,0)$ corresponds to the thermostat rest frame.
One sees that the two integrals, which determine $\overline{p}_{\perp}$,
are both frame dependent due to the presence of $dp_{\parallel}$. However,
the dependence cancels out in the ratio of the integrals. Analogously
one shows that $\Phi_{p_{\perp}}$ is invariant.

We consider again the case of massless particles with vanishing chemical 
potential. As previously the calculations are performed in the thermostat 
rest frame. Then,
$$
\overline{p}_{\perp} = {\pi^5 \over 120 \zeta(3)} {7/6 \choose 1}\, T^3
\cong {2.48 \choose 2.12} \, T \;,
$$
$$
\overline{z^2_{p_{\perp}}} = {8 \zeta(5) \over \zeta(3)}{5/4 \choose 1}\, T^2
- \overline{p}_{\perp}^2 \cong {2.50 \choose 2.40} \, T^2 \;,
$$
$$
{\langle Z_{p_{\perp}}^2 \rangle \over \langle N \rangle }
={1 \over \rho}\Bigg[{4\pi^2 \over 45}{7/8 \choose 1}\,T^5 
- {3 \zeta(3) \over 2\pi}{3/4 \choose 1}\overline{p}_{\perp}\,T^4
+{1 \over 6}{1/2 \choose 1}\overline{p}_{\perp}^2\,T^3 \Bigg]
\cong {2.33 \choose 3.37}\,T^2 \;,
$$
which provide
$$
\Phi_{p_{\perp}} \cong {-0.05 \choose \;\;\;0.29} \,T \;.
$$

When the gas particles are massive and/or the chemical potential
is finite, the correlation measure $\Phi_{p_{\perp}}$ can be numerically 
computed directly from eqs.~(\ref{phipt1}) and (\ref{phipt2}).
In Figs.~1 and 2 we show  $\Phi_{p_{\perp}}$ as function of $T$ and $\mu$
for the pion gas. The pions are, of course, massive with $m=140$ MeV. 
One sees that the presence of the finite mass reduces the correlation 
measure $\Phi_{p_{\perp}}$ when compared to the massless case. 

As already mentioned $\Phi_{p_{\perp}}$ has been experimentally measured 
in the central Pb--Pb collisions by the NA49 collaboration. The result is 
$\Phi_{p_{\perp}} = 0.7 \pm 0.5 $ MeV \cite{Rol97}. If we identify the
system freeze-out temperature with the slope parameter deduced from the 
pion transverse momentum distribution $T \cong 180$ MeV \cite{App97}. 
Then, the value of $\Phi_{p_{\perp}}$, which is read out from Fig.~1 for 
$T=180$ MeV and $\mu=0$, equals 15 MeV and drastically exceeds the 
experimental value. The temperature is significantly reduced if the 
transverse hydrodynamic expansion is taken into account. The freeze-out 
temperature obtained by means of the simultaneous analysis of the single 
particle spectra and Bose-Einstein correlations is about 120 MeV 
\cite{App97}. The value of $\Phi_{p_{\perp}}$ for $T=120$ MeV and $\mu=0$ 
equals 6.5 MeV and is still much larger than the experimental value. 
Let us discuss this puzzling discrepancy.

$\Phi_{p_{\perp}}$ has been measured for pions which come form the limited 
phase-space region: $0.005 < p_T < 1.5$ GeV and $4 < y < 5.5$ \cite{Rol97}. 
However, it should not distort the value of $\Phi_{p_{\perp}}$ noticeably.
First of all, one sees that the acceptance domain of $p_T$ covers the 
$p_T-$region which contributes to the integrals from  eqs. (\ref{phipt1}) 
and (\ref{phipt2}). One notes that the average $p_T$ approximately equals 
$2T$. Secondly, we observe that the system longitudinal expansion influence 
the value of $\Phi_{p_{\perp}}$ insignificantly as long as the transverse 
momentum distribution weakly depends on the particle rapidities. Finally, 
one notes that the size of the longitudinal momentum domain does not
matter very much for the value of $\Phi_{p_{\perp}}$. Therefore, we conclude
that the finite acceptance of the NA49 measurement cannot be responsible 
for the discussed discrepancy.

We have considered three other ways to reconcile the experimental and 
theoretical values of $\Phi_{p_{\perp}}$.
\begin{itemize}

\item The pions are out of chemical equilibrium and the chemical potential
is negative. It appears however that $\mu$ must acquire an unrealistically
large negative value ($\mu = -$ 245 MeV) to get $\Phi_{p_{\perp}}=0.7$ MeV.

\item A substantial fraction of the final state pions come from the hadron 
resonances. These pions do not `feel' the Bose-Einstein statistics at 
freeze-out and should be treated as particles which obey the Boltzmann 
statistics. Then, they do not contribute to $\Phi_{p_{\perp}}$. Assuming 
that the fraction $k$ of the final state pions come from the resonances, 
$\Phi_{p_{\perp}}$ is reduced approximately by a factor $\sqrt{1 - k}$. 
The value of $k$ must be again unrealistically large to reduce 
$\Phi_{p_{\perp}}$ sufficiently.

\item The Coulomb repulsion among the like-sign pions is known to
significantly diminish the bosonic correlations, see e.g. \cite{Bay96}.
However, taking into account the electromagnetic interaction should
not change the value of $\Phi_{p_{\perp}}$ noticeably. The point is
that the Coulomb repulsion moderates the effect of boson statistics
but the attraction among the unlike-sign pions generates the
positive correlation of the similar size \cite{Bay96}.

\end{itemize}

It is possible that the combination of the three effects considered 
above sufficiently reduces the theoretical value of $\Phi_{p_{\perp}}$.
However, it is not a simple problem to perform a numerically reliable 
analysis. Therefore, we leave it for the future studies.

At the end let us comment on a somewhat paradoxical implication of
our study. When the correlation measure $\Phi$ was introduced \cite{Gaz92},
we expected that the value of $\Phi$ would be smaller in A-A than in p-p 
interactions. We shared a rather common opinion that the correlations 
observed in p-p case were of a dynamical origin and tended to be 
washed out in A-A by rescatterings. The recently obtained experimental 
data \cite{Rol97,Adl98} comply with our expectation. Indeed, 
$\Phi_{p_{\perp}} = 4.2 \pm 0.5$ MeV from p-p is about 6 times larger 
than that one from the central Pb-Pb collisions at the same energy 158 
${\rm A\cdot GeV}$ \cite{Rol97,Adl98}. However, we have now found that 
the equilibrium value of $\Phi_{p_{\perp}}$, which significantly exceeds 
$\Phi_{p_{\perp}}$ from Pb-Pb, is close to that one from p-p. This implies 
that even in p-p collisions the origin of the correlations is poorly 
understood. Therefore, a complete solution of the puzzle raised by our 
equilibrium calculation requires a better understanding of the effects 
which control the value of $\Phi$ at N-N level.

\vspace{1cm}

I am indebted to Marek Ga\' zdzicki for initiating this study and 
numerous fruitful discussions. The suggestion by Gunther Roland to
consider the electromagnetic effects is also gratefully acknowleged.

%\newpage

%\newpage
\vspace{1cm}
\begin{center}
{\bf Figure Captions}
\end{center}
\vspace{0.3cm}

\noindent
{\bf Fig. 1.} 
The correlation measure $\Phi_{p_{\perp}}$ as a function of temperature
$T$ for four values of the chemical potential $\mu$. The most upper line 
is for $\mu = 70$ MeV, the second one for $\mu = 0$ etc.

\vspace{0.5cm}

\noindent
{\bf Fig. 2.} 
The correlation measure $\Phi_{p_{\perp}}$ as a function of chemical 
potential $\mu$ for four values of the temperature $T$. The most upper line 
is for $T = 200$ MeV, the second one for $T = 160$ MeV etc.

\end{document}